\documentclass[aps,amssymb,reprint, twocolumn,superscriptaddress,showpacs]{revtex4-1}
\usepackage[]{fontenc}
\usepackage[latin9]{inputenc}
\usepackage{units}
\usepackage{amstext}
\usepackage{amsmath}
\usepackage{amssymb}
\usepackage{graphicx}

\makeatletter
\@ifundefined{textcolor}{}
{%
 \definecolor{BLACK}{gray}{0}
 \definecolor{WHITE}{gray}{1}
 \definecolor{RED}{rgb}{1,0,0}
 \definecolor{GREEN}{rgb}{0,1,0}
 \definecolor{BLUE}{rgb}{0,0,1}
 \definecolor{CYAN}{cmyk}{1,0,0,0}
 \definecolor{MAGENTA}{cmyk}{0,1,0,0}
 \definecolor{YELLOW}{cmyk}{0,0,1,0}
}

\makeatother

\usepackage[english]{babel}
\usepackage{ragged2e}
\begin{document}

\title{High-Q nested resonator in an actively stabilized optomechanical cavity}

\author{F.M. Buters}
\email{buters@physics.leidenuniv.nl}
\affiliation{Huygens-Kamerlingh Onnes Laboratorium, Universiteit Leiden,
2333 CA Leiden, The Netherlands}
\author{K. Heeck}
\affiliation{Huygens-Kamerlingh Onnes Laboratorium, Universiteit Leiden,
2333 CA Leiden, The Netherlands}
\author{H.J. Eerkens}
\affiliation{Huygens-Kamerlingh Onnes Laboratorium, Universiteit Leiden,
2333 CA Leiden, The Netherlands}
\author{M.J. Weaver}
\affiliation{Department of Physics, University of California, Santa Barbara,
California 93106, USA}
\author{F. Luna}
\affiliation{Department of Physics, University of California, Santa Barbara,
California 93106, USA}
\author{S. de Man}
\affiliation{Huygens-Kamerlingh Onnes Laboratorium, Universiteit Leiden,
2333 CA Leiden, The Netherlands}
\author{D. Bouwmeester}
\affiliation{Huygens-Kamerlingh Onnes Laboratorium, Universiteit Leiden,
2333 CA Leiden, The Netherlands}
\affiliation{Department of Physics, University of California, Santa Barbara,
California 93106, USA}

\date{\today{}}

\begin{abstract}
Experiments involving micro- and nanomechanical resonators need to be carefully designed to reduce mechanical environmental noise. A small scale on-chip approach is to add a resonator to the system as a mechanical low-pass filter. However, the inherent low frequency of the low-pass filter causes the system to be easily excited mechanically. We solve this problem by applying active feedback to the resonator, thereby minimizing the motion with respect the front mirror of an optomechanical cavity. Not only does this method actively stabilize the cavity length, but it also retains the on-chip vibration isolation. 
\end{abstract}

\maketitle
\section{Introduction}
Micro- and nanomechanical systems are widely used for precision measurements across a variety of research topics, see Ref. \cite{ekinci2005nanoelectromechanical} for an overview. In cavity optomechanics for example, electromagnetic radiation is used to perform displacement measurements that beat the standard quantum limit \cite{hertzberg2010back,Suh13062014}. Similarly, in atomic force microscopy (AFM) forces as small as 10$^{-18}$ N are measurable \cite{binnig1986atomic}.  

A common challenge when using mechanical resonators is achieving a high mechanical quality factor (Q-factor) and shielding from vibrational noise. Recently, clamping losses, or coupling to external mechanical modes, has been identified as a major source of loss in mechanical systems \cite{cole2011phonon,wilson2011high,yu2012control}, affecting both the mechanical Q-factor and the amount of vibrational noise entering the system. A solution to this problem is to introduce phononic crystals \cite{alegre2011quasi,yu2014phononic} and low frequency mechanical resonators \cite{serra2012ultralow,harry2010advanced,weaver2015nested} to isolate the device from the environment.

Surrounding the resonator of interest with an additional low frequency resonator has a severe drawback. The additional low frequency resonator itself can be mechanically excited by the environment. The obvious solution would be to fixate the additional resonator, but this will remove the effect of the mechanical low-pass filter. Typically this trade-off is circumvented  by reducing the motion of the resonator using active feedback cooling. 

Active feedback cooling, also called cold damping, uses the real time displacement of the mechanical resonator to apply a suitable feedback signal to an actuator (mechanical, optical or electrical) \cite{milatz1953reduction,hirakawa1977damping,cohadon1999cooling,pinard2000full,metzger2004cavity,kleckner2006sub, poggio2007feedback}. Increasing the gain of the feedback signal to the actuator results in more feedback cooling up to the point where the amplified read-out noise causes the mechanical motion to increase.

In this article we will first demonstrate active feedback cooling of our nested trampoline resonator \cite{weaver2015nested}. Because this resonator is part of an optomechanical cavity, a more elegant approach to solving this problem is possible. We will show how, by actively stabilizing the position of the resonator with respect to the front mirror of the cavity, all length variations of the cavity are eliminated, including the motion of the low frequency resonator while retaining the on-chip vibration isolation.

\section{Experimental details}
For the actuation we make use of the dielectric force. This is convenient as any dielectric body in the presence of an electric field gradient experiences a dielectric force \cite{maxwell1868method}. Recently this method was also used to demonstrate control of a micro- and nanomechanical resonators \cite{schmid2006nonconductive,unterreithmeier2009universal}. 

The energy of a dipole in an electric field is $U=-\mathbf{p} \cdot \mathbf{E}$. The force the dipole experiences is $\mathbf{F}=-\nabla U = (\mathbf{p}\cdot\nabla)\mathbf{E}$.  With the dipole moment $\mathbf{p}=\alpha \mathbf{E}$, this can be written as $\mathbf{F}= (\alpha \mathbf{E} \cdot\nabla)\mathbf{E}$. The strength of the dielectric force depends therefore on the applied electric field, its gradient and the polarizability of the medium. Although the nested resonator is largely made from silicon, which is weakly polarizable, the experiment is carried out in vacuum, so relatively large electric fields can be generated. For example, a back-of-the-envelope calculation using a charged sphere placed $\sim \mu m$ distance behind our sample \cite{Bouwers1941} shows that forces up to $\sim \mu N$ are feasible.  

\begin{figure}
\centering{}\includegraphics[scale=0.3]{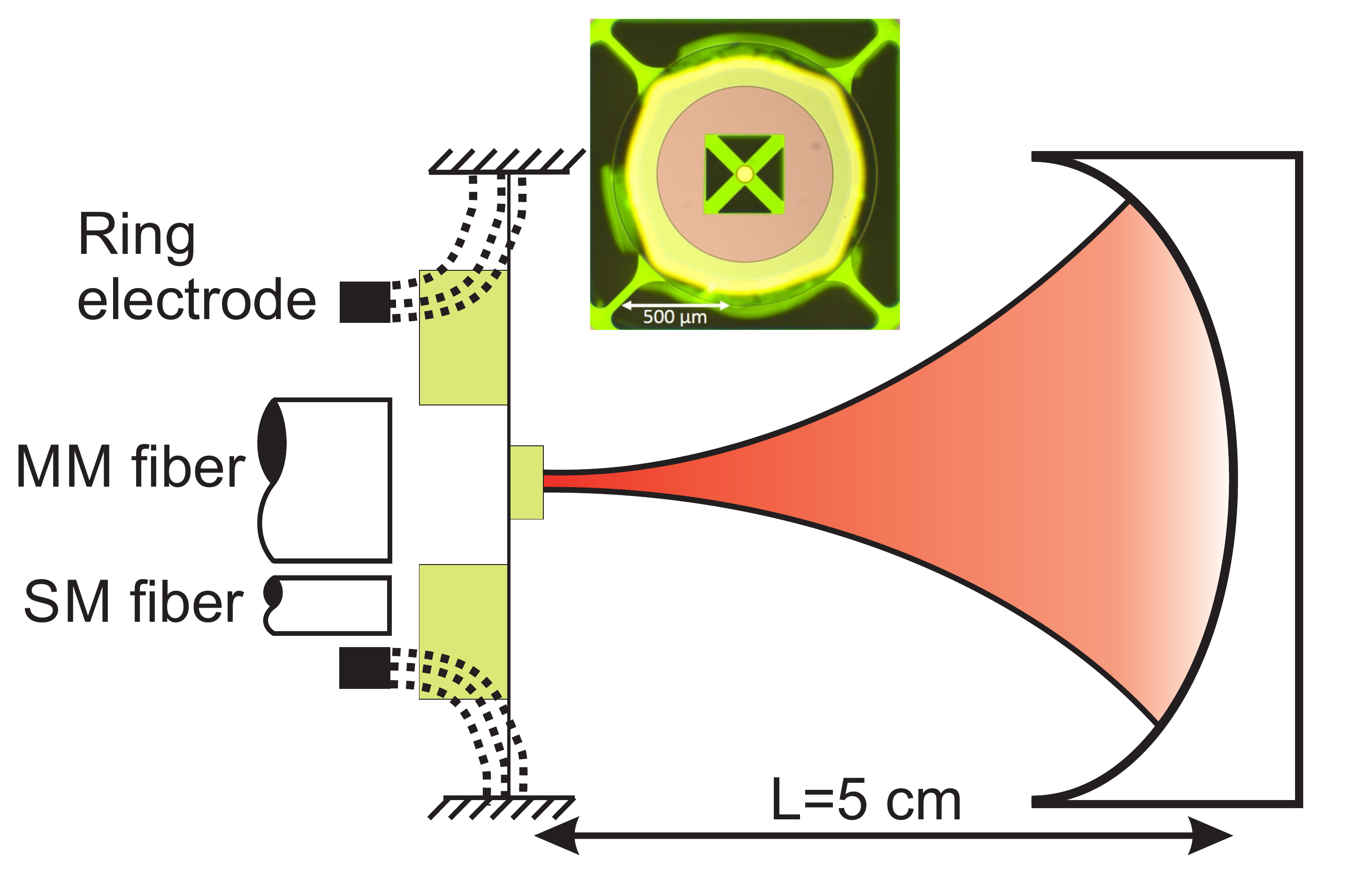}\caption{\label{Electric} Schematic overview of the geometry for applying an electric field gradient. A ring electrode, single-mode (SM) fiber and multi-mode (MM) fiber are placed behind the resonator. The resonator itself is part of an optomechanical Fabry-Perot cavity. Inset: optical image of the nested resonator which consists of an inner trampoline resonator surrounded by a larger outer resonator.}
\end{figure} 

We realize an electric field gradient by placing a small ring electrode behind the nested resonator, as depicted in Fig. \ref{Electric}. The typical distance between electrode and resonator is 500 $\mu m$.  Both a bias voltage and a modulation voltage are applied. Since the electric field is linear in applied voltage, the dielectric force can be written as  $F= \beta V^{2}=\beta(V_{d.c.} + V_{a.c.})^{2} \approx \beta V_{d.c.}^{2}  + 2\beta V_{d.c.}V_{a.c.}$,  in the limit when the applied bias $V_{d.c.}$ is much larger than the modulation $V_{a.c.}$,  with $\beta$ being some constant. Note that the force is now linear in $V_{a.c.}$, therefore a modulation at frequency f$_{0}$ will result in a force at frequency f$_{0}$.

The read-out of the mechanical motion is done in three different ways. A single-mode (SM) fiber is located  approx. 500 $\mu m$ behind the mass of the outer resonator to form a fiber interferometer \cite{rugar1989improved}, as is shown in Fig. \ref{Electric}. The interference signal created by the light reflecting off the outer resonator and the light reflected off the fiber facet allows for a sensitive read-out of the resonators motion. Our fiber interferometer using a standard distributed feedback laser diode at 1550 nm reaches a read-out sensitivity of approx. 300 $\mathrm{fm}/\sqrt{\mathrm{Hz}}$ around 3.5 kHz.

For the second read-out method a low finesse (F=300) Fabry-Perot cavity, 5 cm long operating around 980 nm, is used. The transmitted cavity light is collected using a multi-mode (MM) fiber placed in the center behind the sample as is shown in Fig. \ref{Electric}.

The third method uses the same Fabry-Perot cavity together with a wavelength of 1064 nm to create a high finesse (F$\approx$ 17000) cavity. The Pound-Drever-Hall method \cite{black2001introduction} is used to read out the displacement of this cavity from the light reflecting off  the cavity. Finally, the whole set-up resides in a vibration isolated vacuum chamber with a base pressure of 10$^{-3}$ mbar to eliminate the effect of gas damping on the mechanical properties.

\begin{figure}
\centering{}\includegraphics[scale=0.4]{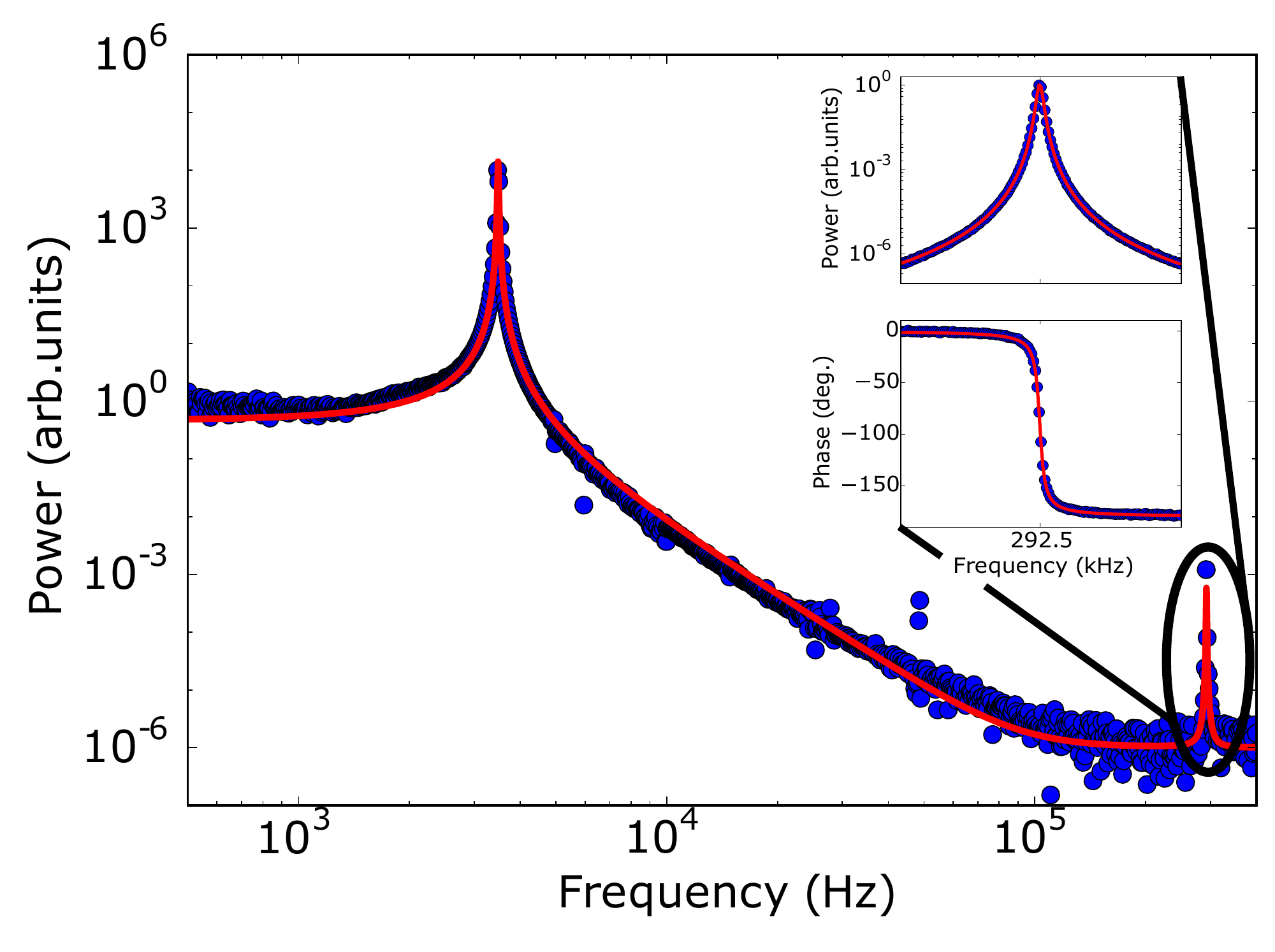}\caption{\label{MechanicalTransfer} The mechanical transfer function of the nested resonator design measured by applying an electrostatic force to the outer resonator and reading out the response of the inner resonator via the cavity. Inset: Driven response of the inner resonator. }
\end{figure}

As a demonstration of the actuation via the dielectric force, the mechanical transfer function of the nested resonator is measured by varying the frequency of the applied force to the outer resonator and reading out the response of the inner resonator via the low finesse optical cavity. The data (blue points) of Fig. \ref{MechanicalTransfer} follow the theory curve for two coupled harmonic oscillators. At the frequency of the inner resonator, which for this sample is 292.5 kHz, more than 60 dB of isolation is provided via the nested resonator design. This is consistent with previous findings \cite{weaver2015nested}. The inset shows that the inner resonator can be driven in this way as well. Note that this measurement assumes a constant force excitation which, judging by the agreement between experiment and theory of Fig. \ref{MechanicalTransfer}, is valid. Therefore the bandwidth of this method of actuation is at least 100 kHz.  

\section{Results}
\subsection{Active feedback cooling}

The main point of the addition of a dielectric force is not to drive the outer resonator, but to reduce its motion. To do so, we use the method of active feedback cooling \cite{milatz1953reduction}. The displacement signal from the fiber interferometer is sent through a differentiating circuit and is amplified with a variable gain amplifier. Together with a DC voltage the signal is then sent to the ring electrode to provide a damping force for the motion of the outer resonator. To avoid difficulty in interpreting the data \cite{poggio2007feedback}, we use an out-of-loop measurement provided by the low finesse cavity to read out the effect of the feedback. 

From the mechanical noise spectrum obtained via the cavity read-out, the effective mechanical Q-factor of the outer resonator is determined together with the total displacement $\left<x^{2}\right>$ by fitting a Lorentzian. Via the equipartition theorem the effective temperature can be calculated using $T = \left<x^{2}\right>M\Omega^{2}_{m}/k_{b}$ with $k_{b}$ the Boltzmann constant, $M = 7\times 10^{-7}$ kg, and $\Omega_{m}=2 \pi \times 3488$ rad/s for this particular sample.

\begin{figure}
\centering{}\includegraphics[scale=0.2]{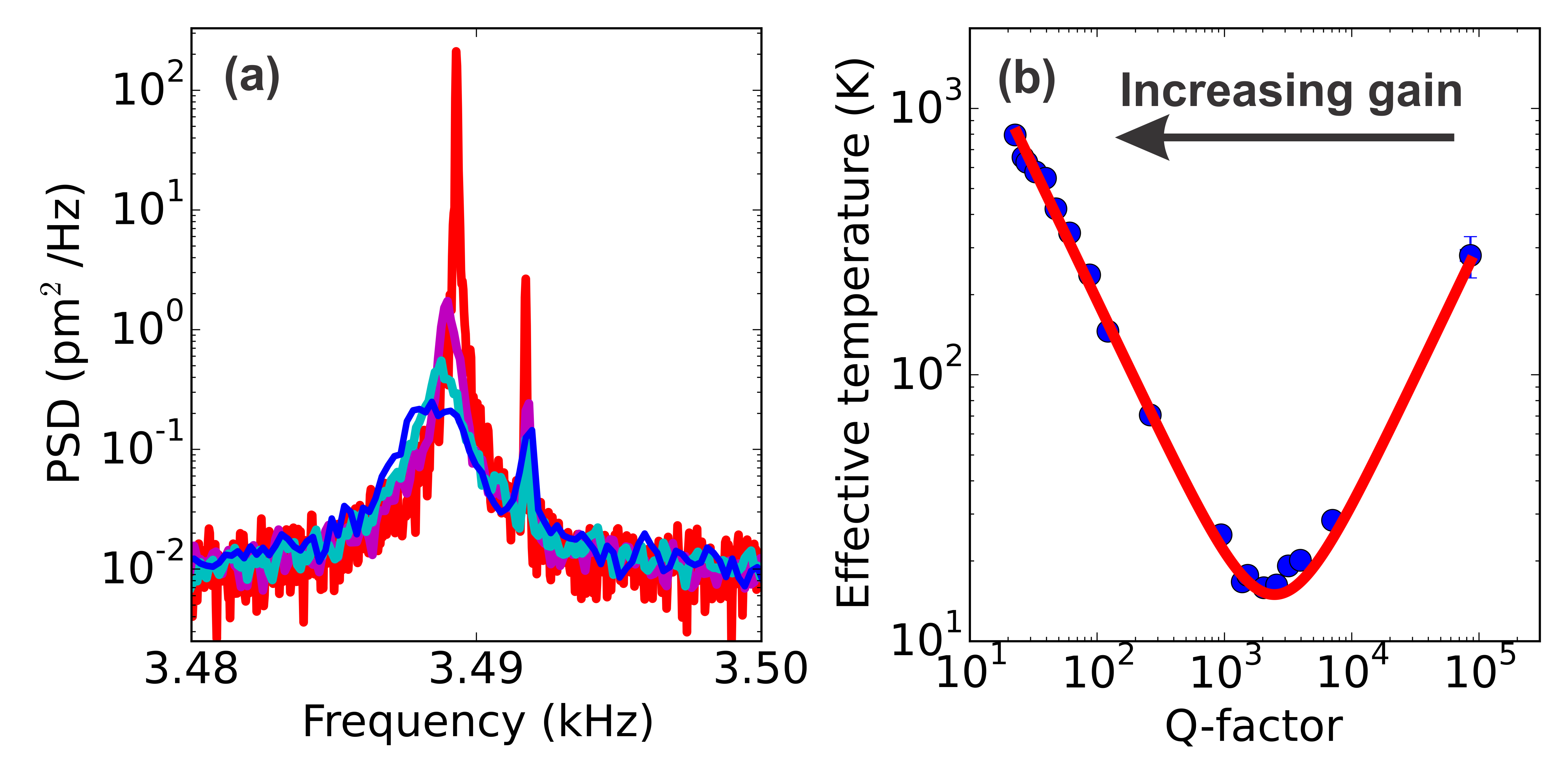}\caption{\label{ActiveFeedback} Active feedback cooling of the outer resonator. When the gain is increased, both the effective temperature and the Q-factor decreases. For large gains, the amplified read-out noise actually drives the outer resonator, increasing the effective temperature. (a) Selection of power spectra demonstrating cooling of the mechanical mode. This out-of-loop measurement uses the read-out via the low finesse cavity. Note that a small additional resonance is visible. (b) Effective mode temperature as a function of mechanical Q.}
\end{figure}
With the feedback circuit activated, the motion of the resonator will be both damped and cooled. Increasing the gain results in more damping and cooling of the mechanical motion up to the point where the amplified detection noise becomes comparable to the thermal noise. The read-out noise from the feedback is transferred to the mechanical motion of the resonator, causing the effective temperature to increase again. We have observed precisely this behavior, as is shown in Fig. \ref{ActiveFeedback}. Note that this is visible because of the out-of-loop measurement. Measuring the mechanical noise spectrum via the interferometer, within the feedback loop, will result in noise squashing \cite{poggio2007feedback}.
 
We are able to reduce the intrinsic Q-factor of 90000 to about 20. The mechanical mode temperature is reduced to about 15 K using this form of cooling. To check if the limiting factor is indeed the noise floor of the interferometric read-out, the data is fitted using the theory derived by Poggio et al. \cite{poggio2007feedback}. For active feedback cooling the effective temperature is given by: 
\begin{equation}
T_{\mathrm{eff}} = \frac{T_{\mathrm{bath}}}{1+g}+\frac{k\Omega_{m}}{4k_{b}Q_{0}}\frac{g^{2}}{1+g}S_{x_{n}}
\end{equation}
with $T_{\mathrm{bath}}$ the environmental temperature, $g$ the feedback gain, $k$ the spring constant, $Q_{0}$ the intrinsic quality factor and $S_{x_{n}}$ the read-out noise power. In our case: $k=340$ N/m, $T_{\mathrm{bath}}=293$ K, and $Q_{0}=90000$. The only free parameter is the read-out noise power $S_{x_{n}}$. From the fit (red curve Fig.\ref{ActiveFeedback}(b)) we obtain a value for $\sqrt{S_{x_{n}}}=380\pm10$ $\mathrm{fm}/\sqrt{\mathrm{Hz}}$, which is close to the noise floor of our interferometric read-out.

Using active feedback the total rms displacement is reduced from 6 pm to 0.8 pm, which for most  applications is sufficient. However in our case, we require further reduction of the motion. To achieve this, the interferometric read-out is replaced with a more sensitive read-out method using a high finesse optical cavity.

\subsection{Active stabilization}
The use of a high finesse optical cavity typically requires a means to keep laser and cavity resonant. Usually, the laser frequency is continuously adjusted to keep it resonant with the cavity mode. An alternative method would be to adjust the cavity length to keep the cavity resonant with the laser. For the set-up depicted in Fig. \ref{Electric}, the cavity length $L$ can be varied by changing the position of the outer resonator using the dielectric force, while tracking the cavity resonance  via the Pound-Drever-Hall (PDH) method \cite{black2001introduction}. In this way the nested resonator design not only mechanically decouples the inner resonator from the environment, but also stabilizes the cavity length with respect to the laser frequency.  

\begin{figure}
\centering{}\includegraphics[scale=0.33]{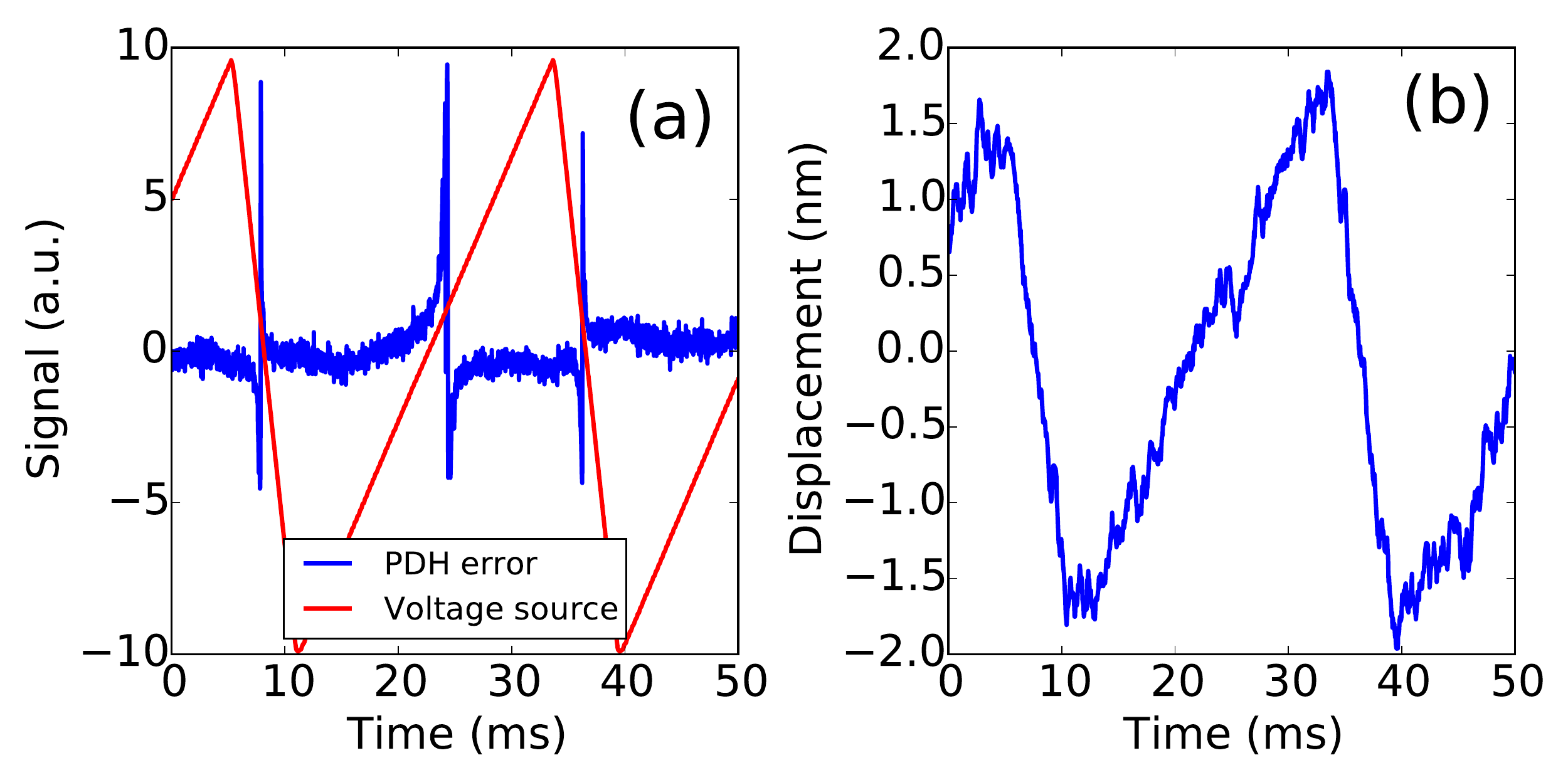}\caption{\label{Openloop} Scanning of the cavity length (a) Varying the position of the outer resonator results in the typical PDH error signal. (b) The displacement of the outer resonator is monitored using the fiber interferometer.}
\end{figure}

In Fig. \ref{Openloop}(a) the typical PDH error signal (blue) is obtained by scanning the cavity length using a high voltage amplifier (red). Note that to ensure a linear response, a DC bias voltage (not shown here) is also added. The displacement of the outer resonator is observed via the interferometer, as is shown in Fig. \ref{Openloop}(b). A displacement of $\pm$1.5 nm provides sufficient range to keep the cavity resonant with the laser. 

We typically require a feedback bandwidth of about 10 kHz to keep laser and cavity resonant. However, the mechanical resonance at  $f= 3488$ Hz provides a very rapid $\pi$ phase shift in the transfer function. A notch filter is placed at $f= 3.4$ kHz in the feedback loop to smooth this transition. A second notch filter at $f=10.2 $ kHz is used to compensate the first higher order mechanical mode.

\begin{figure}
\centering{}\includegraphics[scale=0.33]{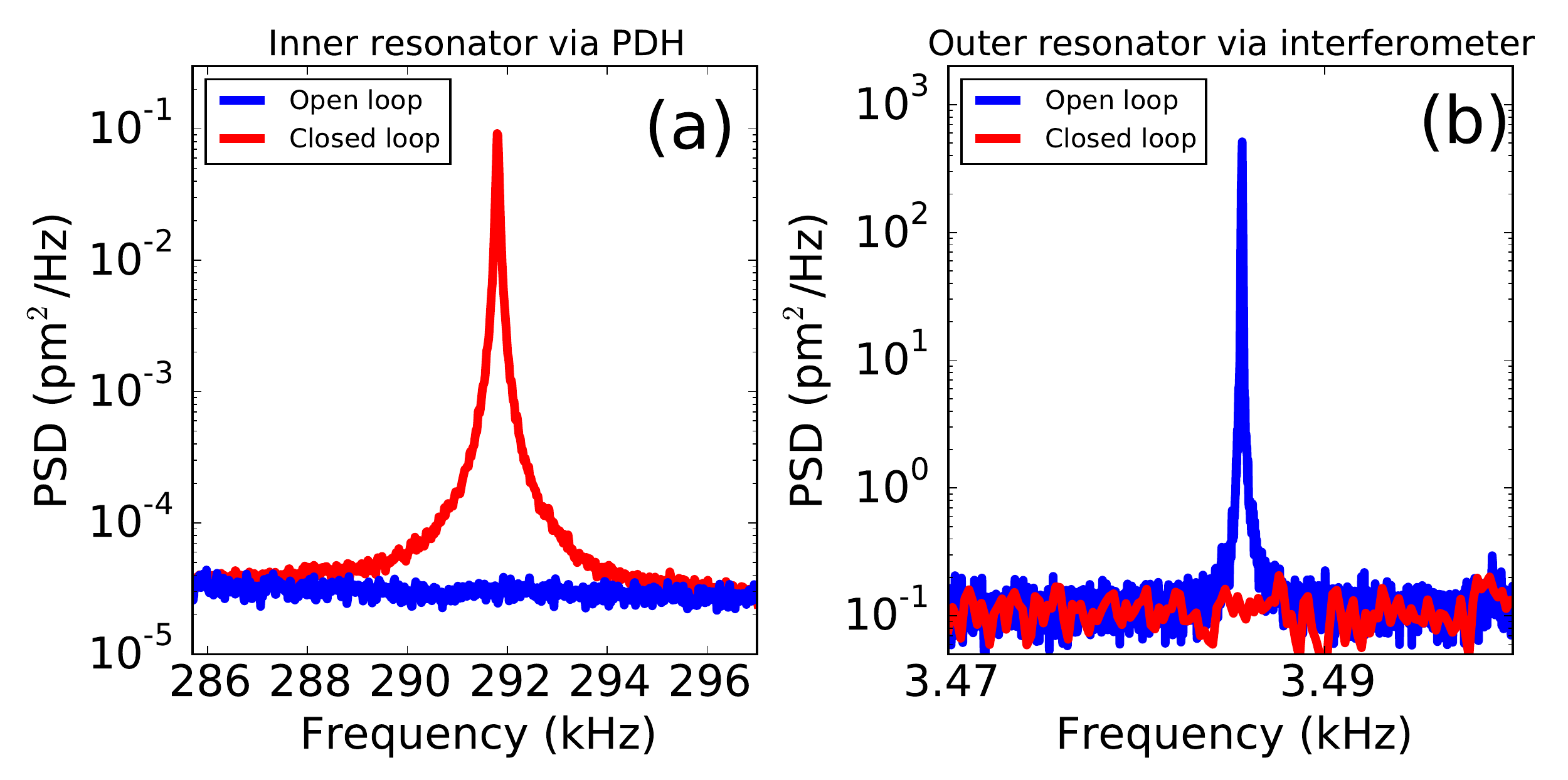}\caption{\label{Closedloop} Active stabilization of the outer resonator. When the feedback loop is closed, the cavity is resonant with the laser, making the motion of the inner resonator visible as is shown in (a). On the other hand, the motion of the outer resonator is no longer visible via the interferometer, as is shown in (b).}
\end{figure}

When the feedback loop is closed, the cavity should remain resonant with the laser. Therefore the motion of the inner resonator, which occurs at a frequency beyond the feedback bandwidth, should be visible in the PDH error signal. Fig. \ref{Closedloop}(a) shows the Fourier transform of this error signal and indeed, with a closed loop, the thermal motion of the inner resonator is visible. The thermal motion of the outer resonator has been mostly eliminated, and is no longer visible via the interferometric read-out, see Fig. \ref{Closedloop}(b). 

From Fig. \ref{Closedloop} it is clear that the feedback not only works, but also that the on-chip isolation still works as evidenced by the clean spectrum of Fig. \ref{Closedloop}(b). However, what has happened to the motion of the outer resonator? Effectively, with active stabilization, a very strong electrical spring is placed between the outer resonator and the front mirror. The only way for the outer resonator to move at a particular frequency, is if the front mirror also moves at this frequency. This stiffening of the outer resonator explains why, with a closed loop, the mechanical motion is no longer visible in Fig. \ref{Closedloop}(b). The additional electrical spring also helps to prevent any unwanted optical spring effects \cite{corbitt2007all} present in an optomechanical cavity.

\section{Conclusion}
In conclusion, we have demonstrated how to solve the problem of fixating an on-chip mechanical low-pass filter while retaining the mechanical isolation. By making use of an optomechanical cavity, the motion of the resonator can be referenced to the front mirror. Not only does this stabilize the cavity with respect to the laser, it also stiffens the resonator, thereby significantly reducing its motion. We have made use of an optomechanical system, but in principle the techniques presented in this work can be applied to any system as long as a suitable reference can be chosen.

\section*{Acknowledgment}
The authors acknowledge the useful discussions with W. Loeffler. The authors would also like to thank H. van der Meer for technical assistance and support.

\bibliography{Optomechanics}

\end{document}